\documentclass[a4paper]{article}
        \newcommand{\be}{\begin{equation}}
        \newcommand{\ee}{\end{equation}}
\usepackage{epsf}

\def\ni{\noindent}
\def\k{\kappa}
\def\r{\rho}
\def\a{\alpha}

\def\G{\Gamma}
\def\d{\delta}

\def\ae{\eta}
\def\z{\zeta}

\def\th{\theta}

\def\m{\mu}
\def\n{\nu}

\def\om{\omega}

\def\l{\lambda}

\def\Z{{\sf Z\kern-4.5pt Z}} 

\def\pa{\partial}
\begin{document}
%\linespread{1.8}
%\doublespacing
%\bibliographystyle{chicago}
\bibliographystyle{unsrt}
%\unitlength=1.00mm
%\special{em:linewidth 0.4pt}
%\linethickness{0.4pt}
%\hoffset=-4.5cm
\author{Alexander~Unzicker\\
        Institute of Medical Psychology\\
        University of  Munich\\[0.6ex]
        Goethestr.~31,  D-80336 M\"unchen, Germany\\
%        Goethestr.~31, \underline{\bf D-80336 M\"unchen}, {\bf Germany}\\
        {\small{\bf e-mail:}  sascha@imp.med.uni-muenchen.de}}

\title{Teleparallel Space-Time with Defects yields
Geometrization of Electrodynamics with quantized Charges} %currents
\maketitle
\large
\normalsize
\begin{abstract}
In  the  present  paper  a geometrization of electrodynamics 
is proposed, which makes use of a generalization of Riemannian geometry
 considered already by Einstein and Cartan in the 20ies. 
Cartan's differential forms description of a teleparallel space--time with 
torsion  %by means of integrability conditions
is modified by introducing distortion 1-forms %besides of 
which correspond to the distortion
tensor in dislocation theory.
%The interplay between the first and second  Bianchi  identity
%and between their structure equations is seen as the action  of an  
%antisymmetry operator (trace of the 
%exterior product with  the basis 1-forms $\om^{\a}$) which is dual to the
% contraction operator. 
Under the condition of teleparallelism, the antisymmetrized part of 
the distorsion 1--form  approximates  the
electromagnetic field, whereas the antisymmetrized part of torsion 
contributes to the electromagnetic current.
Cartan's   structure   equations,  the  Bianchi identities,  Maxwell's
equations and the continuity equation are thus linked in a most simple way.
%The Levi-Civita connection yields
%as usual the gravitational phenomena on a large scale,  whereas
%the teleparallel connection is assumed to describe electrodynamics
%and microphysics.\\ %Thereby, space--time becomes a multiple connected
%manifold with line singularities. \\
After  these  purely geometric considerations a physical interpretation,
 using analogies to the theory of defects in
ordered media, is given.  A  simple defect, which is neither
a dislocation nor disclination proper, appears as source of the 
electromagnetic field. Since this defect is rotational rather than
translational, there seems to be no contradiction to Noether's theorem
as in other theories relating electromagnetism to torsion.
% giving
%evidence for a quantization of the electric charge. 
Then, congruences of defect properties and quantum behaviour
that arise are discussed, 
supporting the hypothesis that elementary particles are 
topological defects of space--time.
%Finally,  the implications on dimensions of the
%respective physical quantities are investigated. 
In agreement with the
differential geometry results, a
dimensional analysis indicates that  the physical unit $(lenght)^2$ rather 
than $As$ is the appropriate unit of the electric charge.\\
%\bf Keywords: geometrization, differential geometry, torsion,
%teleparalleism, homotopy \rm
\end{abstract}

\section{Introduction}
Two independent developments led to the following considerations.
In the early 50ies,
Kondo \cite{Kondo:52} and independently 
Bilby et al. \cite{Bilby:55} discovered 
that  topological  defects  in crystalline bodies, namely dislocations,
have to be  described in terms of differential
geometry.  Cartan's~torsion  tensor was shown to be equivalent to the
dislocation density. It was Kondo himself, who stressed in a series of
papers \cite{Kondo:52} \cite{Kondo:55} that this discovery may have
some  impact  beyond material science.
Kr\"oner  completed  the  theory  in an outstandingly clear way
\cite{Kro:60} \cite{Kro:80} and obtained
many results that remind us from electrodynamics.
% More general texts about dislocations are 
%\cite{Naba:79} and \cite{Mura:87}.
I the meantime, many researchers %\cite{Lago:89} \cite{Mina:71} 
felt particulary attracted
by the beauty of this theory which includes the mathematics of
general relativity (GR) as a special case. 
With Kr\"oner's words: " We  have  seen that
Riemannian   geometry  was  to  narrow  to  describe  dislocations  in
crystals.  Is  there  a reason why space--time has to be described by a
connection  that  is  less general than the general metric--compatible
affine connection ?"\\
The  second reason for dealing with this topic is Einstein's so--called
teleparallelism   attempt  towards  a  unified  field theory,  
grown  out  of  the
correspondence  with  Elie  Cartan \cite{Debe:29} and cumulating in an
article  in  the Annalen  der  Mathematik  in  1930  \cite{Einst:30}. 
Even if this attempt did not succeed,
the fact that Einstein,
trying to create a unified theory of electrodynamics and gravitation,
considered the same extension of Riemannian geometry
 that has shown to describe  defect theory, remains
a remarkable coincidence.
Unfortunately,  most  physicits    have associated
Einstein's  belief in the existence of a unified theory in this
context to his continuous objections to quantum mechanics.
It is one of the main purposes of this paper to show that there is 
 no  contradiction  between  quantum mechanics and a differential
geometry  approach  towards  a  unified  theory. Other interesting
 congruences between quantum behaviour and facts emerging
from geometry have been detected by Vargas \cite{Var:92b} \cite{Var:97b}. 

In section 2, the differential geometry of a 4--dimensional manifold is
revisited  in  differential  forms language using Cartan's moving frame 
method that focusses on integrability conditions. By
introducing   distorsion 1-forms  Maxwell's
equations  appear as purely geometric identities. 
 Thereby  it  is assumed that on a
large  scale,  the  Levi-Civita  connection  describes  as  usual  GR,
whereas  a  teleparallel  connection  governs physics on a microscopic
level,  generating two kinds of geodesics: extremals (depending
on the metric only) and autoparallels.
In section 2.7, some of Einstein's tensor quantities are translated 
into modern forms language. 
Contrarily to Einstein's conviction, 
this  proposal allows singularities in space--time (topological defects). 
In  section  3,  a  visualization  of the obtained results by means of
dislocation theory is given. 
The  starting  point  is  the  Lorentz--invariance  of  the  motion of
dislocations, whereby the velocity of shear waves is analogous to the
velocity of light. 
A generalization of
dislocation theory with finite--size defects, which are described by
homotopy theory, is needed for a
physical interpretation
of  the quantity representing the current. Therefore,
the proposed geometrization  of electrodynamics
seems to require a quantization of the electric charge. 
Further   similarities of defect physics to quantum mechanis are
discussed in section~4,  at  the present stage necessarily in a qualitative manner.
In  section~5, some dimensional analysis remarks about physical units
and the calculability of masses are  given.  Although these remarks may 
not be considered sufficiently
convincing for founding a physical theory by themselves, 
their implications fit  to the previously developed
results.\\
As more recent papers I took inspiration from I should mention
Vargas' papers on geometrization
\cite{Var:91b} \cite{Var:92a}  \cite{Var:93},
Vercin \cite{Ver:90}, who discussed dislocations under the perspective of gauge
theories, and 
Hehl \cite{Hehl:76}, who gave a review of the use of torsion in general
relativity relating the torsion tensor to spin. 
At the end of  section~3, I will discuss the general objections 
formulated by Hehl\cite{Hehl:96}
against relating electromagnetism to torsion and a possible solution.

\section{Differential  geometry  of  a %affinely connected
  4--dimensional  manifold with curvature and torsion}

A scalar--valued $p$--form $\ae$ is called closed, if $d \ae=0$
($d$ is the exterior derivative), and called exact, if a $(p-1)$--form
$\z$ exists with $d \z =\ae$. The rule $dd=0$ implies exact $\Rightarrow$
closed. In other words, if $d \ae$ is a nonvanishing $(p+1)$--form, 
a $\z$ with the above property cannot be defined {\em globally\/}.
  
Regarding the vector-- and tensor--valued forms occurring
in the differential geometry of a 4--dimensional 
manifold the situation is not that simple.
To get an overview, one may list the quantities which are the most important 
ones in the following sense (see in Tab.~I):\\
If the $(p-1)$--forms can be defined globally, 
the respective $(p+1)$--forms have to vanish identically.
These {\em integrability conditions\/} cannot be expressed
by applying a differential operator as when dealing with scalar-valued
forms. Furthermore, the
integrability conditions link the vector--valued with the 
tensor--valued forms.\\

\begin{tabular}{|lcccr|}
\hline
p-form   & symbols &quantity & satisfies &\\
\hline
0-form   & $A^{\ \n}_{\m}, A^{\n}$ &Lie group & \ &\\
1-form & $\om_{\m}^{\ \n},\om^{\n}$ & connection & structure equation&\\
2-form & $R_{\m}^{\ \n}, T^{\n}$ & field & Bianchi identity&\\
3-form & $J_{\m}^{\ \n}, J^{\m}$ & current & continuity equation&\\
4-form & \ & invariant & d=0& \\
\hline
\end{tabular}\\

Table~I.

\subsection{Integrability conditions}
Cartan \cite{Cart:22a} \cite{Cart:22b} developed the theory of 
affine connections 
starting from integrability conditions. 
He introduced the vector equation for a point ${\bf P}$ in an 
arbitrary fixed basis ${\bf a_{\m}}$ as
\be
{\bf P} = {\bf P}_0 + A^{\m} {\bf a_{\m}},
\label{aff1}
\ee
whereas a frame is given by
\be
{\bf e_{\m}}= A_{\m}^{\ \n}  {\bf a_{\n}}.
\label{aff2}
\ee
By differentiating   the   affine  group  with  elements  $(A^{\m}, 
 A^{\ \n}_{\m})$ one obtains the pair $(\om^{\m},\om_{\n}^{\ \m})$,
 the usual exterior derivative operator 
$d$ is here applied also to the basis ${\bf e_{\m}}$ \footnote{This is 
sometimes called {\em exterior covariant\/} derivative.}:
\be
d{\bf P} = d ({\bf P}_0 + A^{\m} {\bf a_{\m}}) =\om^{\m} {\bf e}_{\m}
\label{daff1}
\ee
and
\be
d{\bf e_{\n}}= d (A_{\m}^{\ \n}  {\bf a_{\n}})=\om_{\n}^{\ \m} {\bf e_{\m}}.
\label{daff2}
\ee
$\om_{\n}^{\ \m}$ stands for  $B_{\k}^{\ \m} d A_{\n}^{\ \k}$ and
$\om^{\m}$  for  $ B_{\k}^{\ \m} d A^{\k}$, the matrix $B$ is the inverse
of $A$.
If we ask ourselves, whether the system (\ref{daff1}-\ref{daff2}) %yyy p.6
is integrable, the rule $dd=0$ yields the neccessary conditions, the
Maurer-Cartan structure equations of a Lie group:
\be
d\om^{\l}- \om^{\n}  \wedge  \om^{\  \l}_{\n}=0,
\label{mc1}
\ee
and 
\be
d\ \om_{\l}^{\ \k}
- \om_{\l}^{\ \n} \wedge \om^{\ \k}_{\n}=0.
\label{mc2}
\ee
These are the {\em integrability conditions\/} for the system
 (\ref{daff1}-\ref{daff2}),
the neccessary conditions for manifold to be locally affine space,
thus the neccessary conditions for defining $A_{\m}^{\ \n}$ and
$A^{\n}$ {\em globally\/}.
On the other hand, eqns.~(\ref{mc1}) and (\ref{mc2}) will be used
as definitions for torsion and curvature if the terms on the
r.h.s. do not vanish. Torsion and curvature stand for the failure
of integrability of the system (\ref{daff1}) and (\ref{daff2}).\\

\subsection{Equations of structure and Bianchi identities}

Going one level down in Tab.~I, analougous arguments
apply: rather than  integrating the connections and
obtaining the Lie group, the connections are now  considered
as the basic quantities. For example, in GR, the Riemannian curvature
tensor  is defined by
\be
R_{\n \m \l}^{.\ .\ .\ \k}=
\pa_{\n} \G_{\m  \l}^{\k}-
\pa_{\m} \G_{\n  \l}^{\k}+
 \G^{\k}_{\n \r} \G^{\r}_{\m \l}-
 \G^{\k}_{\m \r} \G^{\r}_{\n  \l},
\label{4.1}
\ee
\ni
where  $\G_{\m  \l}^{\k}$ is the Levi-Civita-connection.
However, eqn.~(\ref{4.1}) holds  as well for a more
general  affine  connection which is not neccessarily symmetric in the
lower two indices and not completely determined by the metric.
In  differential  forms  language, eqn.~(\ref{4.1}) is called the {\em
second structure equation\/} and takes the form
\be
R_{\l}^{\ \k}=d \om_{\l}^{\ \k}
- \om_{\l}^{\ \r} \wedge \om^{\ \k}_{\r},
\label{2str}
\ee
using the antisymmetric properties of the exterior algebra
 and omitting  the  form  indices in eqn.~(\ref{4.1}). %One should note that
%(\ref{2str})  is  {\em  not\/}  obtained  by  applying  the  covariant
%derivative with respect to all 3 indices of the
%connection.  This would be a innatural operation that does not lead to
%integrability  conditions. 
Using differential forms language of Cartan
puts in evidence the fundamental difference between the value indices
$\l$ and $ \k$ and the form indices $\m$ and $\n$ (eqn.~\ref{4.1}).
 The latter define 
the surface on which a 2--form `lives'.
$R_{\l}^{\ \k}$ is the  curvature 2-form and
$  \om_{\l}^{\  \k}$ is the connection 1-form, both of them take values
in the Lie algebra of the affine group.
%$d$  denotes again the exterior derivative.
To obtain the integrability conditions for the connections, one has to
differentiate (\ref{2str}):
\be
d\ R_{\l}^{\ \k}-\om^{\ \k}_{\n} \wedge
 R_{\l}^{\ \n} -  \om^{\ \n}_{\l} \wedge  R^{\ \k}_{\n}=0,
\label{2bia}
\ee
This is called the  {\em  Second  Bianchi identity}.
Analogously to (\ref{mc1}) with nonvanishing r.h.s.,
 one differentiatiates the vector-valued basis
1-forms $\om^{\l}$, and  obtains the torsion   tensor:
\be  T^{\l}= d\om^{\l}- \om^{\n}  \wedge  \om^{\  \l}_{\n} ,
\label{1str}
\ee
which  is  called  {\em  first}  structure
equation. The integrability conditions for the basis 1--forms
are obtained by  differentiation of the vector-valued
2-form torsion: 
\be
d\ T^{\k}+ T^{\n} \wedge
\om^{\ \k}_{\n} -  \om^{\n} \wedge  R^{\ \k}_{\n}=0,
\label{1bia}
\ee
which is called  {\em  First Bianchi identity}. 
The Bianchi identities are the integrability
conditions the fields have to satisfy in order to yield well--defined
connections. If  torsion and curvature are chosen  independently 
without satisfying the Bianchi identities,
the pair of connections $(\om^{\m},\om_{\n}^{\ \m})$ cannot be 
defined any more.
In the following, we will investigate the interplay of the two branches
that led to the 1st and second Bianchi identity.
In a situation with vanishing curvature, i.e. breaking only the
integrability conditions for the vector equation, one can still
integrate eqn.~\ref{mc2} and obtain a globally defined frame
 ${\bf e}_{\m}$, whereas the 
opposite constraint, curvature with zero torsion like in GR,
does not even allow the integration of eqn.~\ref{mc1}, because
the nonintegrable frame ${\bf e}_{\m}$ `spoils' also 
eqn.~\ref{mc1}.\\
In conclusion, one may, discarding the connections,
descend further in Tab.~I and
consider the integrability conditions for the fields:
the currents $J_{\m}^{\ \n}$ and $J^{\n}$ , 
now defined as the nonvanishing  r.h.s. of
(\ref{2bia}) and (\ref{1bia}) \footnote{See also Hehl \cite{Hehl:91}.}
still have to satisfy their continuity 
equations (vanishing 4-forms in Tab.~I) in order to yield 
well--defined fields. Since this extension is not neccessary for the following,
I will not go into details here.

\subsection{Teleparallel description of General Relativity}

In the following we restrict to a metric-compatibile connection
($\omega_{\mu \nu}+\omega_{\nu \mu}=0$). Then, I will consider a 
teleparallel geometry,
that means the  curvature 2--form $R_{\l}^{\ \m}$ vanishes everywhere.
This does not inhibit a geometric description of the energy-momentum
tensor, rather it  can  be  seen  as  a  formal  replacement  of the
Levi--Civita  connection by a teleparallel connetion. Since there is a
freedom  in choosing the connection this can always be done by adding
to $\G_{\m \n}^{\k}$ the so--called contorsion tensor \cite{Schouten:54} \cite{Hehl:71}
\be
S_{\m \n}^{\ \ \k}:= T_{\m \n}^{\ \ \k}- T_{\n \ \m}^{\  \k}
+T_{\ \m \n}^{\k},
\label{cont}
\ee
where  $T_{\m \n}^{\ \ \k}$ are the components of torsion.
In  this  case  the  Riemannian  curvature  tensor  of  GR (which is
obtained  from  the  Levi--Civita connection) can be expressed in terms
of torsion and its derivatives (cfr. \cite{Schouten:54}, 4.22).
In this case, the usual geodesics (extremals determined by the metric
only) have to be distinguished from autoparallels.\\
\begin{figure}[hbt]
\epsfxsize=7.0cm
\epsffile{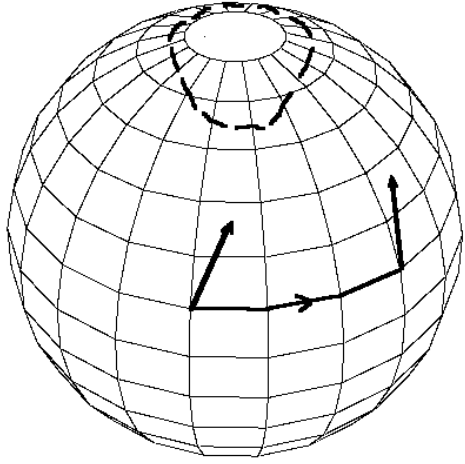}
\caption{If autoparallels on the unit sphere are defined according
to the Mercator projection, one obtains a teleparallel structure
with vanishing curvature. The parallel transported vector changes direction
in the imbedding 3-dimensional space, whereas the angle to the meridians is 
kept fixed.}
\label{sphere}
\end{figure}
To illucidate the interplay curvature--torsion, the simple example
Fig.~\ref{sphere} (cfr.\cite{Naka:95}, sec.7.3) is given.
The ususal geodesics (extremals) on a sphere are great 
circles, but one may alternatively define parallel transport by keeping
the angle between the vector  and the straight line in Mercator's
projection fixed. The geometry becomes teleparallel, but has 
nonzero torsion. The meridians are in this case autoparallels 
rather than geodesics determined by the metric (extremals). 
Transporting a vector along a closed path 
around one of the poles (see dotted line in Fig.~\ref{sphere}), 
however, yields
a whole turn of $2 \pi$. We may regard the two poles as having
Dirac-valued curvature. In this case, integrating the curvature
over the whole sphere still yields $4 \pi$, as required
by the Gauss-Bonnet theorem.\\
These topological issues were addressed first by Cartan in his letter
to Einstein from Jan 3, 1930 \cite{Debe:29}:
`Every solution of the system (\ref{mc2}) creates, from the topological 
point of view, the continuum in which it exists' \footnote{See also the
comments given by Vargas \cite{Var:97a}.}.
%In a similar manner, one may change the Levi-Civita connection that
%describes GR and obtain teleparallelism all over  space--time.
%The price one has to pay is singularities (point or line defects)
%where the torsion is divergent. As a consequence, space--time becomes
%multiple connected.
%On a large scale, however, when one integrates the (Dirac-valued) 
%singularities, space--time still has curvature and  the Levi--Civita 
%connection describes macrogeometry properly. For a teleparallel version
%of GR, see also \cite{Hehl:91}. 

\subsection{The distortion 1--forms $\th^{\m}$}

In conventional tensor analysis, traces are important because they 
are   invariant  under coordinate transformations. The same holds for the 
antisymmetric part of a tensor. In differential forms
language, the latter corresponds to exterior multiplication with  
the basis 1-forms $\om^{\m}$, whereby  the  sum  over  
the doubled index $\m$ is taken\footnote{Analogously, contraction is performed by {\em interior\/} 
multiplication with the basis 1--forms $\om^{\m}$. Contraction and 
antisymmetrizing are dual.}. This  
antisymmetry operator ${\cal A}$ that raises the degree of a
form,  but  lowers  the  degree  of  the  value  indices,
whereas the exterior derivative $d$ raises the degree of a
form,    without   changing   the   type   of   the   form   (tensor-,
vector-valuedness). \\
If we investigate the equations of 2.2, we may visualize the respective
contributions of these operators to the quantities in Tab.~I 
in the following sketch:

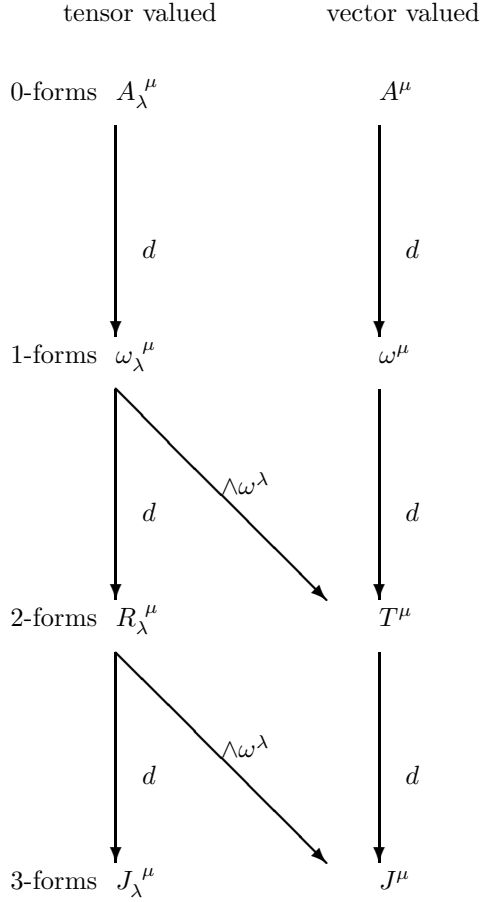
\begin{figure}[hbt]
\setlength{\unitlength}{0.7mm}
\begin{picture}(77,180)(-25,-5)
\thicklines
\put(0,150){\vector(0,-1){40}}
\put(50,150){\vector(0,-1){40}}
\put(0,155){${A_{\l}^{\ \m}}$}
\put(50,155){${A^{\m}}$}
\put(0,100){\vector(0,-1){40}}
\put(0,100){\vector(1,-1){40}}
\put(50,100){\vector(0,-1){40}}
\put(0,50){\vector(0,-1){40}}
\put(0,50){\vector(1,-1){40}}
\put(50,50){\vector(0,-1){40}}
\put(0,105){${\om_{\l}^{\ \m}}$}
\put(0,55){${R_{\l}^{ \ \m}}$}
%\put(0,5){${\om_{\l}^{\ \k} \wedge R_{\k}^{\ \m}}$}
\put(0,5){${J_{\l}^{\ \m}}$}
\put(50,55){${T^{\m}}$}
%\put(50,5){${R_{\l}^{\ \m} \wedge \om^{\l}}$}
\put(50,5){${J^{\m}}$}
\put(5,25){${d}$}
\put(5,75){${d}$}
\put(5,125){${d}$}
\put(55,25){${d}$}
\put(55,75){${d}$}
\put(55,125){${d}$}
\put(-20,5){3-forms}
\put(-20,55){2-forms}
\put(-20,105){1-forms}
\put(-20,155){0-forms}
%\put(-10,170){$SO_0(3,1)$ valued}
\put(-10,170){tensor valued}
\put(40,170){vector valued}
\put(50,105){${\om^{\m}}$}
\put(20,80){${\wedge \om^{\l}}$}
\put(20,30){${\wedge \om^{\l}}$}
\label{p1}
\end{picture}
\caption{From top to bottom, the degree of the respective differential
forms increase; from left to right, the number of value indices decreases.
The $p$-forms exist only, if the respective $(p+2)$-form vanishes.}
\end{figure}

By the action of ${\cal A}$
the connection 1--form is transformed into a vector valued 2--form
(which is,  in the holonomic case, torsion);
the tensor--valued Riemannian curvature 2--form is  transformed into
a vector--valued 3--form $\om^{\n} \wedge R^{\ \k}_{\n}$ which 
contributes to the `current' of torsion $J^{\m}$.
We may formally extend this action of the antisymmetry operator
to the `top level' of the left column in Fig.~2,
the 0--form $A_{\m}^{\ \n}$ 
which takes values in the linear group, and consider besides 
$\om^{\n}$ the term
\be
\th^{\n}:= \om^{\n} -  \om^{\m} \wedge  A_{\m}^{\ \n} 
\label{def}
\ee
or briefly $\omega^{\nu} - d A^{\nu}$,
which I shall call {\em distortion\/} 1--form, referring already
to the physical interpretation given in section~3.
In Euclidean space, one could write\footnote{When dealing with 0-forms, 
one may omit the wedge.}
\be
\th^{\n} = B_{\k}^{\ \n} d A^{\k} -A_{\m}^{\ \n} \wedge 
B_{\k}^{\ \m} d A^{\k}=(B_{\k}^{\ \n}-\d_{\k}^{\ \n}) d A^{\k},
\label{kro}
\ee
where $\d_{\m}^{\ \n}$ is the Kronecker delta (0--form).
It follows easily that $d \th^{\n}=d \om^{\n}$. Therefore,
the first  Bianchi identity is not 
affected if in the first structure equation $d \om^{\l}$ is replaced by
$d \th^{\l}$. 
This definition takes into account that  $d A_{\m}^{\ \n}$ and $d A^{\m}$
are different quantities that should consequently 
be `transformed back' also by different quantities   $B_{\n}^{\ \m}$ and
$(B_{\m}^{\ \n}-\d_{\m}^{\ \n})$.
Cartan frequently called  $(\om^{\m},\om_{\n}^{\ \m})$ the pair of 
connections. In a certain sense it is more justified to call
$(\th^{\m},\om_{\n}^{\ \m})$ the pair of connections since 
both $\th^{\m}$ and $\om_{\n}^{\ \m}$   can be brought 
to zero {\em locally\/}  by an appropriate Lorentz
transformation.
%if the manifold is simple--connected and Euclidean
%($T^{\m} and R_{\m}^{\ \n}$ vanish).

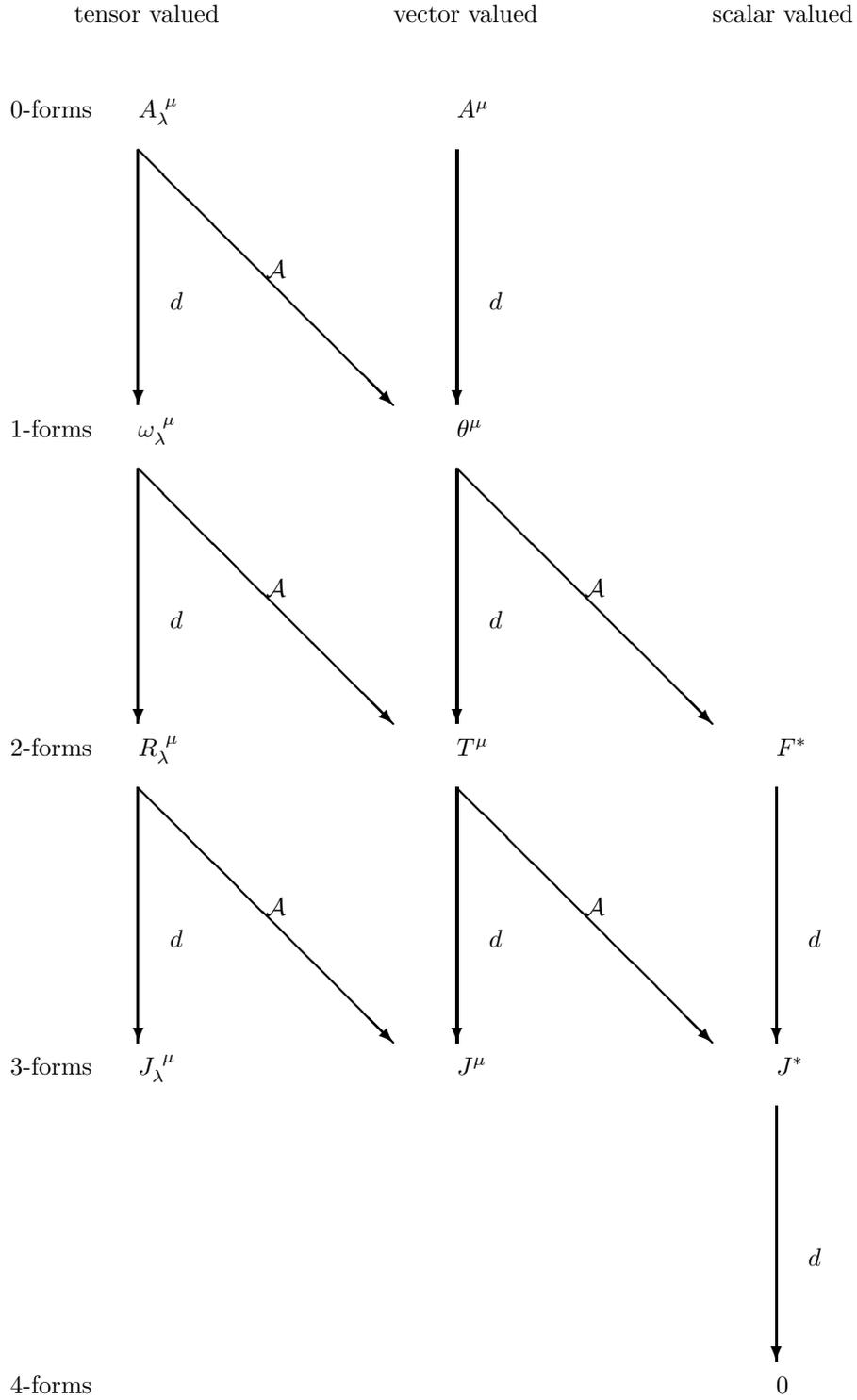
\begin{figure}[hbtp]
\setlength{\unitlength}{0.9mm}
\begin{picture}(145,220)(-25,-50)
\thicklines
\put(0,150){\vector(0,-1){40}}
\put(50,150){\vector(0,-1){40}}
\put(0,155){${A_{\l}^{\ \m}}$}
\put(50,155){${A^{\m}}$}
\put(0,100){\vector(0,-1){40}}
\put(0,100){\vector(1,-1){40}}
\put(50,100){\vector(0,-1){40}}
\put(0,50){\vector(0,-1){40}}
\put(0,50){\vector(1,-1){40}}
\put(50,50){\vector(0,-1){40}}
\put(0,105){${\om_{\l}^{\ \m}}$}
\put(0,55){${R_{\l}^{ \ \m}}$}
%\put(0,5){${\om_{\l}^{\ \k} \wedge R_{\k}^{\ \m}}$}
\put(0,5){${J_{\l}^{\ \m}}$}
\put(50,55){${T^{\m}}$}
%\put(50,5){${R_{\l}^{\ \m} \wedge \om^{\l}}$}
\put(50,5){${J^{\m}}$}
\put(5,25){${d}$}
\put(5,75){${d}$}
\put(5,125){${d}$}
\put(55,25){${d}$}
\put(55,75){${d}$}
\put(55,125){${d}$}
\put(-20,5){3-forms}
\put(-20,55){2-forms}
\put(-20,105){1-forms}
\put(-20,155){0-forms}
%\put(-10,170){$SO_0(3,1)$ valued}
\put(-10,170){tensor valued}
\put(40,170){vector valued}
\put(0,150){\vector(0,-1){40}}
\put(0,150){\vector(1,-1){40}}
\put(0,100){\vector(0,-1){40}}
%\put(50,150){\vector(1,-1){40}}
\put(50,100){\vector(1,-1){40}}
\put(50,50){\vector(1,-1){40}}
%\put(50,0){\vector(1,-1){40}}
\put(100,50){\vector(0,-1){40}}
\put(100,0){\vector(0,-1){40}}
%\put(100,100){\vector(0,-1){40}}
%\put(130,10){\vector(-2,3){27}}
%\put(100,105){${u}$}
\put(50,105){${\th^{\m}}$}
\put(100,55){${F^{*}}$}
\put(100,5){${J^{*}}$}
%\put(100,-45){${T^{\m} \wedge T_{\m}}$}
\put(100,-45){${0}$}
%\put(130,5){${A{*}}$}
\put(105,-25){${d}$}
\put(105,25){${d}$}
%\put(105,75){${d}$}
%\put(115,25){${\delta}$}
\put(20,80){${\cal A }$}
\put(20,30){${\cal A }$}
\put(20,130){${\cal A }$}
%\put(70,130){${\cal A }$}
\put(70,80){${\cal A}$}
\put(70,30){${\cal A }$}
%\put(70,-20){${\cal A }$}
\put(-20,-45){4-forms}
\put(90,170){scalar valued}
\label{p2}
\end{picture}
\caption{The basis 1-forms $\om^{\mu}$ in Fig.~2 are replaced by
 the deformation
 1-forms $\th^{\mu}$. The extension of the relations in Fig.~2 leads to
Maxwell's 2nd pair of equations as Bianchi identity in the `0th' 
column to the right.}
\end{figure}

\subsection{Maxwell's equations} 
\label{Max}
In the most general situation  outlined in  Fig.~3, I consider 
again  a teleparallel geometry
with a vanishing   curvature 2--form $R_{\l}^{\ \m}$, that means
the connection $\om_{\m}^{\ \n}$ is integrable and the 0-forms
$A_{\m}^{\ \n}$ can be defined in every point of the manifold.\\
A straightforward extension of the 
relations in Fig.~2 is applying the antisymmetry operator
to the  distortion  1-forms  $\th^{\m}$ (Since
$\om^{\m} \wedge  \om_{\m}$ is zero, this would have been senseless
without introducing  $\th^{\n}$), which is equivalent to
 applying the antisymmetry operator
with respect to both indices \footnote{To extend the action of ${\cal A}$  to
covariant  indices,  one  has  to  lower  the  index  of $\om^{\n}$ by
multiplication with the metric: $\om_{\m}:=g_{\m \n}\ \om^{\n}$.} 
of $A_{\m}^{\ \n}$:
\be
F^{*}:=-\om_{\n}  \wedge \th^{\n}  
=A_{\m}^{\ \n} \wedge \om_{\n} \wedge \om^{\m} .
\label{em}
\ee
For reasons that will become clear soon I call the resulting 2-form
$F^{*}$, the tensor
{\em dual\/} to the electromagnetic field.
\footnote{The Hodge star operator $*$ is an isomorphism between $p$-forms
and $(4-p)$-forms. We assume the  Jacobian determinant as 1.}
%\footnote{Einstein made
%in his 1930 paper \cite{Einst:30} a similar proposal while dealing 
%with  the `antisymmetric part $a$' of the vierbeins $h$ in 
%section 5 `first approximation'.}.
Exterior differentiation yields
\be
J^{*}:=d F^{*}=d \th^{\m} \wedge \om_{\m} -\th^{\m} \wedge d \om_{\m} .
\label{dF}
\ee
with the 3-form $J^{*}$. Analogous to the
 other relations in Fig.~3, the antisymmetrized torsion 
$T^{\m} \wedge \om_{\m} $  contributes to $J^{*}$ \footnote{It
is worth mentioning
that the quantity $T^{\m} \wedge \om_{\m}$ was considered already by
Einstein \cite{Einst:30}, eqn. 33 and \cite{Einst:30a}, eqns.
(31)-(32). In the context of Chern--Simons theory, \cite{Hehl:91}, 
\cite{Miel:92} and \cite{Cha:97} discussed it.}.\\
%Considering $d \om^{\n} =d \th^{\n}$ and the definition of $\om^{\n}$
%and  $\th^{\n}$, (\ref{dF}) may also be written as
%\be
%J^{*}=d F^{*}=d \om_{\m} \wedge d A^{\m}.
%\label{dF2}
%\ee
As the reader may have noted, one can now obtain Maxwell's 2nd pair 
of equations $\delta F := *d* F=J$ by identifying the 2-form $F^{*}$ 
with the  {\em dual\/} of the electromagnetic field 2-form $F$ and by 
identifying $J^{*}$ with the 3--form dual to the current $J$.
Poincare's lemma $dd=0$, applied to $F^{*}$, yields the continuity
equation. Both   equations   appear  in the   `0th column' to the right
 of Fig.~3 as Bianchi identity and continuity equation
\footnote{Hehl \cite{Hehl:76} calls  an  identity involving the tensor of
nonmetricity `0th' Bianchi identity.}.\\
Eqn.~\ref{dF},  written as $\d F = J$ does not
determine completely the 2--form $F^{*}$. The remaining degree of
freedom can be used to satisfy Maxwell's 1st pair of equations, 
$dF=0$, or equivalently, by introducing the vector potential $A$
with $dA=F$. One should not forget, however, that due to deRham's 
theorem, there is still a degree of freedom left for $F$, since
every harmonic \footnote{More precisely, one should say {\em 
primitively\/} harmonic form, since we may deal with nontrivial 
topologies.} form $H$ satisfies $dH=\d H=0$. Therefore, $F$ is 
only determined up to a harmonic form.\\

\subsection{Nonlinearity}

The $A_{\mu}^{\ \nu}$ are elements of $GL(4)$.
If we consider the subgroup $GL(3)$,
 antisymmetrizing  the elements
 of $GL(3)$ with 
${A_{\m}^{\ \n} \wedge \om_{\n} \wedge \om^{\m}}$ corresponds
(up to a double cover) to a projection on $SO(3)$ and a linearization.
The $A_{\mu}^{\ \nu}$ give  information 
about the distorsion (dilatation and shear) and 
orientation of a volume element
with respect to a given coordinate system, the 
${A_{\m}^{\ \n} \wedge \om_{\n} \wedge \om^{\m}}$ about 
the orientation only. \\
$SO(3)$ is a deformation retract of the nonsingular elements of
$GL(3)$. Applying the 
antisymmetry operator
with respect to both indices means 
projecting from $GL(3)$ to $SO(3)$,
with the restriction that the resulting  term appears in the 
vesture of a 2-form which can be written as an antisymmetric
matrix.
Thus in first approximation,  multiplication of $SO(3)$-matrices
can be done by adding their antisymmetric parts, that means,
in first approximation,  one may describe 
the electromagnetic field as a 2-form, and in first 
approximation, the superposition principle holds.

\subsection{Relations to the Einstein-Cartan TP attempt}

The above considerations on
$A_{\mu}^{\ \nu} \wedge \omega_{\nu} \wedge \omega^{\mu}$
were inspired by Einstein's 1930 paper.
 I will explain which of Einstein's tensor quantities
correspond to forms in the above sections, referring to equation
numbers there \cite{Einst:30}.

Einsteins vierbeins $h_s^{\nu}$ (section~2) correspond to  
$A_{\mu}^{\ \nu}$. 
Eqn.(12), though in tensor language, is equivalent to the definition 
of the $\omega_{\mu}^{\ \nu}$ I repeated in section~2.1
(he writes both $h$ for the Matrix and its inverse).
I should say here that I do not propose Einsteins (29) and (30),
together with their definitions (27) and (28),
as  field equations. (27) may be seen as interior covariant
derivative of torsion (cfr.~\cite{Var:97c}) but (28) does not define 
a reasonable quantity from the differential forms perspective.

In his section `first approximation', Einstein considers
the  quantities $\bar h_s^{\nu}$ defined in (37).
To translate this into forms language, I 
introduced the  $\theta^{\nu}$'s in section~2.4. 
The  $\theta^{\nu}$'s, however, are not neccessarily 
small as $\bar h_s^{\nu}$ is small compared to 
$h_s^{\nu}$.
 If we go ahead, Einstein considers the antisymmetric part
of the $\bar h_{a \mu}$, $\bar a_{a \mu}$ (eqn.~45).
Since the only possible `translation' of  $\bar a_{a \mu}$ is 
$A_{\mu}^{\ \nu} \wedge \omega_{\nu} \wedge \omega^{\mu}$,
Einstein's `electromagnetic field'  $\bar a_{a \mu}$
coincides with the quantity $F^{*}$ I proposed as dual to 
the electromagnetic field.

\section{An elastic continuum with defects as model for a space--time
with particles}

Differential geometry has shown to describe the physics of defects
\cite{Kondo:52} \cite{Bilby:55} \cite{Kro:60} \cite{Kro:80}.
Cartan's structure equations and the Bianchi identities are the 
natural nonlinear generalizations of the definitions and the governing
equations in defect theory \cite{Ver:90}. For example, the first Bianchi
identity expresses the fact that dislocations may not end inside a 
crystalline body (teleparallelism).
I will now  use dislocation theory for visualizing the above
results. Since in a dislocated
crystal directions of vectors
may be compared globally, it can be described by a  teleparallel geometry 
discussed in the previous section. It is clear that the physics of defects 
in a real crystal cannot be completely equivalent to the physics of space--time,
 but one may use the concept of the `continuized crystal' \cite{Kro:80}
it as a model providing further insight. %I will, however, not only
%assume finite dislocation densities as in the limiting process 
%pointed out  by \cite{Kro:80} which
%justifies the description of dislocation {\em density\/} 
%as torsion 2-form, but also consider line singularities which are
%Dirac--valued in torsion.
%In  this  case  one would have a space-time-continuum in which
%curvature   vanishes  everywhere but due to   line
%singularities  it  becomes a multiple connected manifold.\\
It will be helpful here to be familiar with the  
concept of the `internal observer' in a crystal introduced by
Kr\"oner \cite{Kro:80}, see also \cite{Mis:90}. The internal observer 
measures distances by counting 
lattice points. He is unable to detect deformations
or waves of the elastic space-time-continuum, as long those do not
manifest themselves in defects. 
%This reminds us from a quantum mechanical observer, 
%who's measuring process requires the interaction of elementary particles.
The most important presupposition for a spacetime-analogy, however,  
is the appropriate description of Lorentz--invariance.

\subsection{Lorentz-invariance in dislocation theory}

 %They may, however, end in 
%disclinations \cite{deWit:77a}, which are described by the Riemannian
%curvature tensor, which was shown by
%Kr\"oner \cite{Kro:91}, who
%completed  the description of defects (dislocations and disclinations)
%in terms of differential geometry.\\
%However, despite various attempts \cite{Edelen:80}
%\cite{Kos:77a} \cite{Kos:77b} \cite{Ant:74}  a  consistent  theory  of defect dynamics is still
%missing \cite{Kro:94}. An excellent overview  gives\cite{Kro:80}.
The discovery of a  relativistic  behaviour  of  dislocations
goes back to  Frank \cite{Frank:49} and Eshelby 
\cite{Eshelby:49} in 1949.
They showed that  when a screw dislocation moves with velocity
$v$ it suffers a longitudinal contraction by the factor 
$\sqrt{1-\frac{v^2}{c^2}}$, where $c$ is the velocity of transverse
sound. The total energy of the moving dislocation is given by the
formula %$E=E_0 \sqrt{1-\frac{v^2}{c^2}}$,
$E=E_0/ \sqrt{1-\frac{v^2}{c^2}}$,
 where $E_0$
is the potential energy of the dislocation at rest.
These old, but exciting results  were recently extended
 \cite{Gun:88} \cite{Gun:94} to a 
Lorentz-invariant theory of defect dynamics.
In real media, two velocities for longitudinal and transversal
sound exist. This was considered as an 
obstruction by several authors  \cite{Hehl:94}
\cite{Gun:96} to a complete analogy between a continuum with defects and
space-time with matter, since longitudinal sound is always faster and two 
different $c$'s would `destroy' the relativistic description. However,
 space--time can be assumed to be `incompressible'\footnote{To my surprise,
 something similar has been already proposed in 1839 by MacCullagh 
\cite{MacC:39}. In fact, his theory of the rotationally elastic
aether, who's equivalence to Maxwell's equations in vacuo is known
for 158 years now, corresponds in first order to the interpretation
of the electromagnetic field given in section~\ref{Max}.}. %Since every particle
%(defect) causes also shear distorsions of the continuum, there is no
%problem with relativistic effects due to two different sound velocities.
If one goes to the limit of infinite velocity for longitudinal sound, the
formulas (12) and (13) in \cite{Eshelby:49} yield only
distorsions of shear type.
Since every defect causes also shear distorsions,  it causes shear
distorions only in the limit of incompressibility. Therefore
no defect defect matter may propagate faster %\footnote{G\"unther 
%\cite{Gun:67} investigated tachyonic phenomena in crystals. I cannot
%discuss this here.} 
than the
velocity of transverse sound, otherwise its energy would become
infinite.\\
Since space--time is no ordinary matter, there is no physical contradiction
in the assumption of incompressibility.
Therefore, following \cite{Ver:90}, defect dynamics may be described 
formally in a 4--dimensional space--time with torsion and Lorentzian 
signature of the metric. \\

%\subsection{`Quantization' of screw dislocations}
\subsection{The most simple defect -- an electron ?}

There are two distinct types of dislocations, screw and edge
dislocations, each of them causing
 different distorsions of
the crystal. From this follows that there is a certain
{\em separability\/} of the physics of screw and edge 
dislocations. Of particular interest are here screw dislocations,
because the expression obtained above by antisymmetrizing the
torsion tensor gave a contribution to the current  $J^{*}$.
Torsion is equivalent  to the dislocation density tensor 
$\a_{\k \m}^{\ \ \l}$ and
the  density of screw dislocations is described by
mixing the  indices $\k \m \l$, this is sometimes
called {\em H-torsion\/}.

Before going further in relating the two types of dislocations
to  electromagnetism and gravitation, one has to realize that
dislocations are {\em line\/} singularities, whereas elementary 
particles are expected to be point-like defects. 
Therefore, we are interested in finding the most simple possible
defects in an elastic continuum that (at least macroscopically)
appear as point-like. Since dislocations cannot end within the crystal
unless there is curvature, it is an immediate guess  to consider 
closed dislocation loops.
The problem is that in crystals no closed screw dislocation
loops exist. Rather
closed  loops  of dislocations consist of two pairs of screw and edge
dislocations of opposite sign each other \footnote{closed {\em edge\/}
dislocation loops instead may exist, see \cite{Kru:60} for a discussion.}.
 This  defect can be
visualized by cutting the continuum along a surface,
displacing the two faces against each other by the amount of the Burgers
vector and rejoining them  again  by gluing. \\
Similarly we can think of cutting a (circular) surface,
{\em twisting\/} the faces, and gluing them together  again
(see Fig.~4).
This {\em would\/} correspond  to a closed screw dislocation loop, but
 a crystal lattice
resembling distant parallelism cannot be defined any more.
\begin{figure}
\epsfxsize=9.0cm
\epsffile{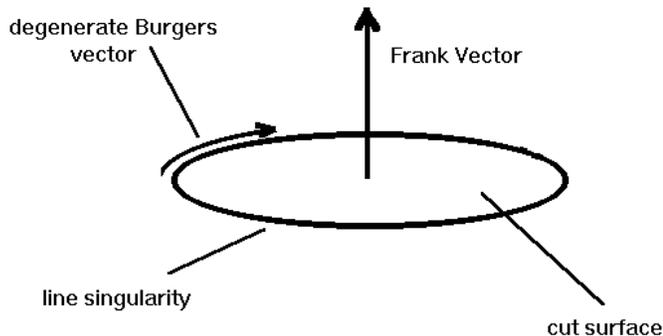}
\caption{A screw dislocation has a Burgers vector parallel to
the dislocation line. The defect depicted here is locally equivalent
to a screw dislocation, if the twisting  (Frank) vector is small, 
otherwise the Burgers vector degenerates to a curved object.}
\label{sdl}
\end{figure}
In another context, this  kind of defect has been investigated by
Huang and Mura \cite{Mura:70}, who called it {\em edge disclination\/},
referring  to disclination
theory. The twisting angle is called {\em Frank vector\/} there.
If a  vector is transported parallely along a path going
through  this  `screw  dislocation  loop',  the  twisting  would  yield a
nonvanishing  Riemannian  curvature  tensor  (which  indeed, describes
defects in disclination theory).\\
In the case the `screw dislocation loop' is a Dirac--valued line 
singularity of finite size, one can resolve, however,
the problem by allowing only {\em multiples of\/} $2 \pi$ as twisting angles,
thus maintaining the teleparallel structure. 
Precisely this defect has been described by Rogula \cite{Rog:76}, who consideres it
a `third' type of defect which is neither a dislocation nor a disclination.
For the following reasons I consider this defect a good candidate for
 describing the electron:
\begin{itemize}
\item
On the large scale, its defect density becomes approximated by the
(antisymmetrized) H-torsion, which gives a contribution to the current
$J^{*}$ in section~\ref{Max}.
\item
The  defect is  a source of the deformation 
that corresponds to the electric field.
\item
Two versions of this defect exist, which according to its screwsense,
could represent an electron or positron. This `handedness'
of the defect would explain CP-violation.
\item
H-torsion %, which describes  the defect density on a large scale,
couples to spin $1/2$-fields \cite{Obu:82} \cite{Obu:83a} \cite{Obu:83b}.

\end{itemize}

\subsection{Description by means of homotopy theory}

Topological defecs are classified by homotopy groups. If we look at
the static case of  three dimensions, 
line  defects are described by the fundamental group $ \pi_1$, 
whereas the second homotopy group $\pi_2$ classifies point defects.\\
In  terms of principal fibre bundles, during parallel transport along a 
path a vector undergoes a transformation, which is  in the above case 
an element
of the fibre $SO(3)$. The closed path through the `screw dislocation loop' 
yields a whole rotation by an angle of $2 \pi$,
corresponding  to a nontrivial element  of  the  first  homotopy group of
the fiber $SO(3)$. Since $\pi_1(SO(3))$ is $\Z_2$, we face the problem that
{\em two\/} defects, each of them representing an electron, cancel 
out\footnote{The same holds for  the Lorentz
group, $SO(3,1)$.} by the rule $1+1=0$. 
Coiled line defects, however, do influence also
higher homotopy groups (for example $\pi_2$ is influenced by $\pi_1$ of
the projective plane, see \cite{Naka:95} \cite{Min:80}).\\
Since in the case of $SO(3)$ $ \pi_2$ is $0$, no topologically 
stable point defects may exist\footnote{This is in agreement with the fact 
there are no elementary particles with radial symmetry.}. The solution to this dilemma 
may be a defect  called Shankar's monopole \cite{Shan:77}, representing the the
nontrivial element $1$  of the {\em third\/}
homotopy group $\pi_3(SO(3))$, which is \Z.\\
Considering  their `screwsense', 
it is impossible  that two  `screw dislocation loops'
 could merge in a way
that makes the distorsion of the continuum
disintegrate completely. I suppose rather that two `screw dislocation loops'
form a Shankar monopole. In this case, $\pi_1$  vanishes, but 
 $\pi_3$ (rather than $\pi_2$) would be 
influenced by the fundamental group $\pi_1$.\\

\subsection{Some implications and objections}

Given the approximation in section~\ref{Max}, 
the  0th component of the  3-form H-torsion
 is proportional to the amount of area enclosed in the `screw 
dislocation loop',
since the length of the dislocation  -- assuming multiples of $2 \pi$
as Frank vector -- is multiplied with a `degenerate' Burgers vector, whose
length is again proportional to the length of the loop. Therefore, charge
can be seen  as  the  amount  of  `twisted  area' of all defects in a 
volume, regardless their directions of the Frank vector.
To ease understanding, 
only the static 3-dimensional case of screw dislocations was discussed here,
which corresponds to the 0th component of the 4-dimensional current
(charge). One should, however, remember, that the Lorentz-invariant properties
of defect dynamics allow a formal description in 4~dimensions.
 Therefore components involving time should behave alike.

The electromagnetic field, according to this proposal, takes values in
the  
Lorentz group, $SO(3,1)$, a pure electric
field in the subgroup $SO(3)$. % comment on no mag. monopoes  deswegen?
This sounds very strange, since the {\em entire\/} electromagnetic field,
 not only the purely electric or magnetic
part could vanish under Lorentz transformations.
This does, however, hold only {\em locally\/}. If we consider 
of the `closed screw dislocation loop' which, according to its
screwsense, represents an electron or positron, its inside is rotated 
by a amount of $\pi$ relative to a point at infinite distance where the 
electric field vanishes. A rotation of the coordinate system could make
the inside and vicinity of the electron nearly field--free but would
cause a homogenous electric field of (maximum) value $\pi$ far from the 
defect. Therefore, such a transformation changes not only the electromagnetic
field but also the nature of its  test particles in a manner that leaves
the physical situation % equations of motion
unchanged. In other words, the topology of a space--time with these
defects generates a preferential coordinate system, according to which
we usually define the electromagnetic field.

A serious objection against theories relating 
torsion to electromagnetism is the following:
Torsion is related to translations
and translations are related to energy-momentum via Noether's
theorem, `and nothing else', as Hehl \cite{Hehl:96} states.
In the present proposal, the electromagnetic field is related
to the quantity ${A_{\m}^{\ \n} \wedge \om_{\n} \wedge \om^{\m}}$
(cfr.~sec.~2), which does {\em not\/} contain torsion.
I suggested, however, that the antisymmetrized torsion 
${T^{\n} \wedge \om_{\n}}$ contributes to the electromagnetic
{\em current\/}. Being a $3$-form, it is not reasonable to
integrate this quantity over a surface, as one does with the 
torsion $2$-form which yields then a translation.

Furthermore, H-torsion is only an approximation for the defect density.
The `closed screw dislocation' defect  proposed as
elementary particle of the current, is, as Rogula \cite{Rog:76} 
explains, a defect of its own type. From the arguments in section~3.2
it is obvious that it is a {\em rotational\/}
defect rather than a translational one. Therefore, Noether's
theorem seems not to contradict this proposal. On the other hand,
if torsion can serve as an approximation only, a precise
differential geometric descrition of the above defect is desirable.

\section{Quantum behaviour of defects}

It is interesting that
the restriction of teleparallelism, that led to Maxwell's equations
in section 2.6, applied to dislocation theory, led to a 
quantization of the term $T^{\n} \wedge \om_{\n}$ which contributes
to the electric charge.

Topological defects, however, share most interesting properties with
the quantum behaviour of particles. Firstly, a sign change in the 
homotopic classification
of a defect describes an `antidefect', corresponding to
 the phenomenon of 
every particle having an antiparticle.
This allows an obvious and intuitive understanding 
of the pair creation and pair annihilation processes.

\begin{figure}[hbt]
\epsfxsize=9.0cm
\epsffile{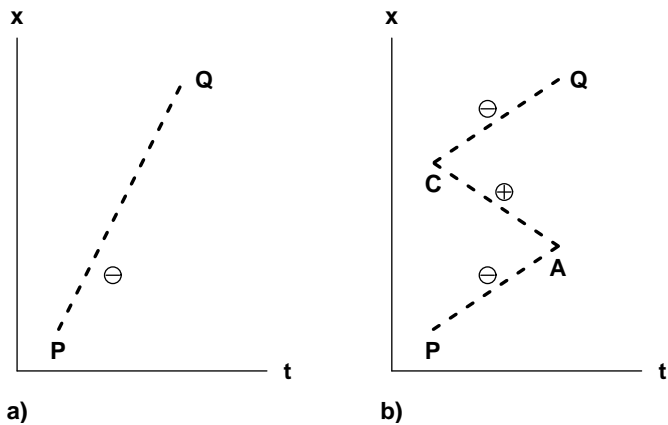}
\caption{The propagation of dislocations, `screw disloaction loops', 
or electrons is completely analogous. By measuring the events in $P$ and $Q$,
there is no method to
detect whether a defect propagating from  $P$ to $Q$ goes a `direct'
path (a) or has a creation--anihilation process plugged in between (b). 
The `Feynman diagrams' (a) and (b) are indistinguishable.}
\label{feynm}
\end{figure}

Fig.~\ref{feynm} a) shows how the motion of a single dislocation in a crystal
from Point $P$ to $Q$ is indistinguishable from a process that involves
an anihilation of two dislocations of opposite sign in $A$ and  a creation
of two dislocations in $C$. 
Analagously, if we interpret the defects 
in Fig.~\ref{feynm} as `screw dislocation loop' and its inverse 
(electron and positron), it can be seen as Feynman 
Diagram with two an extra couplings (an additional virtual photon 
travels from $A$ to $C$ backwards in time).
 
If one measures only the `departure' of an electron in $P$ and the 
`arrival' in $Q$, it is clear that it makes no sense to speak about
a trajectory of an elementary particle.  Considering the double slit 
experiment it makes no sense to say the defect went the one way or the 
other. This famous consequence 
of the quantum mechanics does not appear mysterious any more.

Then, topological defects themselves are as indistinguishable,
if their homotopic classification coincides. As elementary
particles,  one cannot describe them with classical statistics.

In such a space--time, only defects are detectable. I refer here again to the
concept of the `internal observer' in a crystal introduced by
Kr\"oner \cite{Kro:80}. Any quantum mechanical observer is an `internal
observer' in this sense. He may by no means detect distorsions
or waves of the elastic space-time-continuum, as long those do not
manifest themselves in defects. 
Defects cannot be described properly as waves only, nor
being classical particles. Rather a field may be seen as a `tendency
to generate defects'.

If space--time is distortable, one may assume that under large or enduring 
stress, it `wrenches' and builds defect pairs.
The tendency to produce these topological defects should be governed by the 
value of Planck's constant $h$.
Regarding determinism, a `background temperature'
consisting of oscillations of the elastic continuum may cause non 
predictable random fluctuations of the equations of motion on a 
microscopic level (vacuum fluctuations).
Thus complete determinism would be impossible
as a matter of principle.

\section{Dimensional analysis}
Dimensional  analysis,the method on which the following remarks
 are based, has been  developed by  Bridgeman \cite{Bridgeman:31}.
Recent
work in analysing unification theories by considering fundamental
constants was done by \cite{Treder:92}, \cite{Ross:89},
\cite{Hantzsche:90} and  \cite{Biswas:91}.

\subsection{Definitions}
There is an  analogy  between   vectors  in  a
n-dimensional vector space and fundamental constants. $n$ vectors $v_i$
are linear independent if
\begin{equation}
\sum_{i=1}^{n}  a_i v_i=0
\end{equation}
implies $a_i=0$ for all $i$.
Let's call $dim(c)$ the SI units of an expression $c$ containing
 fundamental constants. The Operator $dim$ defines an equivalence
 relation, for example $n~\cong~1$ holds for any real number $n$.\\
%In   the   following,   we  will  not  distinguish  any  more  between
%$dim(c)$ and $c$. \\
$n$ fundamental constants $c_i$ are called independent, if from
\begin{equation}
dim(\prod_{i=0}^{n} c_i^{\gamma_i})=1,
\end{equation}
$d_i \in {\rm R}$, follows $\gamma_i=0$ for all $i$.
For example, the speed of light $c$, dielectricity $\epsilon$ and
permeability %yyy
 $\mu$ are not independent because
$\mu \epsilon$= $\frac{1}{c^2}$.
A set of fundamental constants generates a space of SI units; from
$h$,  $G$  and  $c$ we obtain by a `basis transformation` the SI units
$m$, $kg$, $s$:
\begin{equation}
\left( \begin{array}{cc}
m\\
s\\
kg\\
\end{array} \right)
\cong
\left( \begin{array}{c}
-3/2 \quad  1/2 \quad 1/2\\
-5/2 \quad 1/2 \quad 1/2\\
\quad 1/2 \quad 1/2  -1/2
\end{array} \right)
\left( \begin{array}{c}
c \\
h \\
G
\end{array} \right).
\end{equation}\\
where the matrix elements denote exponents. Addition in the common
 matrix algebra is replaced by multiplication.
The thus obtained units are known as Planck's units.
\subsection{The vector space of fundamental constants}
I will prove now :\\
$kg$ $\notin$ span($h$, $c$, $\epsilon$, $e$).\\
Proof:$h$, $c$, $\epsilon$, $e$ are dependent, because
\begin{equation}
dim(\frac{e^2}{h \/  c \/  \epsilon}) \cong 1,
\end{equation}
because the fine   structure  constant $\alpha$ is dimensionless.
It follows span($h$, $c$, $\epsilon$, $e$)= span($h$, $c$, $e$),
because ($h$, $c$, $e$) are independent.
If
\begin{equation}
dim(h^{\alpha} c^{\beta} e^{\gamma} kg^{\delta}) = 1,
\end{equation}
then  $\gamma=0$,  because  there  is  no  way  of  getting rid of the
Amp\`eres.  While  the ligth speed $c$ transforms    $s$ into
 $m$ only, the $m^2$ in the
denominator of dim(h) can never be compensated by any power
$\alpha$, $\beta$ and $\delta$ . Therefore,
$\alpha=\beta=\delta=0$ follows.\\
\bf Given the present unit system, any formula for the electron mass
involves necessarily $G$.\\
\rm
\noindent
This gives some evidence that a unification of electromagnetism
and  quantum  theory  could only be achieved in the context of general
relativity, and therefore differential geometry.
For several reasons, however, I doubt that - holding up the present
physical  unit system - a unified theory that predicts masses could be
obtained {\em at all\/}:
\begin{itemize}
\item
There are basically two possibilities of obtaining mass from the set
$h$, $c$, $\epsilon$, $e$  and $G$:
$\sqrt{\frac{e^2}{\epsilon G}}$ and
$\sqrt{\frac{h c}{G}}$. The first does not contain $h$
and can therefore not
resemble  quantum  behaviour,  whereas  the latter has neither $e$ nor
$\epsilon$,  consequently  no  electrodynamics in it\footnote{This is
true as long as a theoretical  prediction  of  the  fine structure
constant, that may reveal a link between $e$ and $h$,   is  missing.}.
\item
Both expressions  differ by 20 orders of magnitude  from the
electron  mass.  It is  very  unlikely that a unifying theory can
give a simple formula for a factor $10^{20}$. A similar remark was
given in \cite{Bleeker:81}.
\item
It   would   be   still   an  open  question  to  calculate  the  {\em
electromagnetic  part}  of  the  electron  mass  (an expression, that
obviously should not involve $G$).
\end{itemize}
The electromagnetic units Amp\`ere, Volt etc. are rather
arbitrary.  Let  us  remind  that  at Maxwell`s time Coulumb's law was
written   $F=\frac{e^2}{r^2}$,   and  therefore  $dim(e)=\sqrt{\frac{kg
m^3}{s^2}}$ or $m \sqrt{N}$. These conventions obviously do not change
physics  (with  the  old  system one can't calculate masses either, of
course).  It  does  not matter {\em whatever\/} unit one chooses for the
elementary   charge.   Therefore,  without  doing any harm, a `purely
geometric' unit like $m^2$ can be defined as measuring charge.
The units of physical quantities would change as follows:\\

\begin{tabular}{|lcr|}
\hline
Quantity          &  present units  &       new units\\
\hline
Charge   & $As$ & $m^{2}$\\
Current & $A$ & $m^2 s^{-1}$\\
Potential & $V$ & $N m^{-1}$\\
Dielectricity($\epsilon$) & $A s V^{-1} m^{-1}$ & $m^2 N^{-1}$\\
Permeability($\mu$) & $V s A^{-1} m^{-1}$ & $kg m^{-3}$\\
Electric field & $V m^{-1}$  & $N m^{-2}$\\
Magnetic field & $V s m^{-2}$ & $N s m^{-3}$\\
Magnetic flux &V $s$ & $N s m^{-1}$\\
\hline
\end{tabular}\\

Table~2.
\vspace{1.0cm}\\
As one can easily verify, all physical laws remain unchanged.
Of course, the choice of $m^2$ is motivated by the fact that the
antisymmetric part of torsion, like torsion itself, has the 
physical unit $m^{-1}$, or $m^2$ per volume. Looking at Fig.~3, all
quantities on topleft -- bottomright diagonals have the same physical
units.
By modifying the unit system in the proposed manner,
 one gaines
the  possibility of obtaining a formula
of  the  electron  self-energy,  for  example $\frac{h c}{\sqrt{e}}$ -
without $G$. 
If one relates this order of magnitude to the experimental value 
$E_{el}=0.511 MeV$, the
electron  can  be  assumed to be a topological defect (as described in
section 3.2)  of the order
$10^{-24}  m^2$, the square of Compton's wavelength. This is 
certainly
be more realistic than the Planck length of $10^{-35} m$, but can 
hardly be tested unless an experimental method for
determining the size of the topological defect is developed.  If,
however,
other types of defects representing neutrons and protons can be
found, a prediction of mass ratios should be possible if one assumes
that the sizes of the respective defects (in $m$ or $m^2$) have 
simple ratios.

\section{Outlook}

The Lorentz-invariance in defect  dynamics has only be proven
rigidly for straight screw disocations. Although the defects
described here can be expected to behave in the same manner,
a (much more complicated) proof has still to be given.\\
To derive equations of motion, Lagrange densities have to be found.
Until  now,  a  defect  description  has  only  been  proposed for the
electron,  not  for the neutron and the proton. The success or failure
of  the  present  theory  will  depend on the possibility of finding a
model  also  for  the  latter elementary particles. \\
The possibility of calculating self--energies of elementary particles,
however, does not seem remote, since the electromagnetic field, taking 
values in $SO(3,1)$, is finite everywhere. Unfortunately, the lack of 
experimental methods for measuring the `radius' of the electron does not
 allow a testable prediction. Additional models for other particles should,
however, lead to a prediction of the respective mass ratios. Furthermore,
the violation of the superposition principle for electromagnetic fields
may be tested experimentally.\\
The main conceptual advantage of defect theory is that many
 properties  of elementary particles to which we are familiar from
 experiments, like pair
creation  and anihilation, quantum statistics, wave--particle dualism,
antiparticles,  CP  violation,  nonexistence  of  radial  symmetry and
others appear  to have a certain logical interplay.\\

%\newpage
\large
Acknowledgement\\
\normalsize

I am grateful to  Dr.~Karl~Fabian for guiding my attention 
to Prof. Kr\"oner's
work. I owe a lot to Prof.~Jos\`e~G.Vargas for teaching me Cartan's 
moving frame method and for commenting on a first version of
this paper.

\end{document}